\pdfoutput=1
\documentclass[twoside,twocolumn,9pt]{article}

\usepackage{extsizes}
\usepackage[super,sort&compress,comma]{natbib}
\usepackage[version=3]{mhchem}
\usepackage[left=1.5cm, right=1.5cm, top=1.785cm, bottom=2.0cm]{geometry}
\usepackage{balance}
\usepackage{times,mathptmx}
\usepackage{sectsty}
\usepackage{graphicx}
\usepackage{lastpage}
\usepackage[format=plain,justification=raggedright,singlelinecheck=false,font={stretch=1.125,small,sf},labelfont=bf,labelsep=space]{caption}
\usepackage{float}
\usepackage{fancyhdr}
\usepackage{fnpos}
\usepackage[english]{babel}
\usepackage{array}
\usepackage{droidsans}
\usepackage{charter}
\usepackage[T1]{fontenc}
\usepackage[usenames,dvipsnames]{xcolor}
\usepackage{setspace}
\usepackage[compact]{titlesec}
\definecolor{cream}{RGB}{222,217,201}
\usepackage[activate=normal]{pdfcprot}
\usepackage[hidelinks]{hyperref}
\usepackage{amsmath,amsthm,amssymb,amsfonts,latexsym,bbm}

\newcommand{\bfe}{\mathbf{e}}
\newcommand{\bff}{\mathbf{f}}
\newcommand{\bfi}{\mathbf{i}}
\newcommand{\bfj}{\mathbf{j}}

\newcommand{\bfm}{\mathbf{m}}
\newcommand{\bfn}{\mathbf{n}}
\newcommand{\bfo}{\mathbf{o}}

\newcommand{\bfQ}{\mathbf{Q}}

\newcommand{\bfR}{\mathbf{R}}
\newcommand{\bfr}{\mathbf{r}}
\newcommand{\bfu}{\mathbf{u}}
\newcommand{\bfv}{\mathbf{v}}
\newcommand{\bfw}{\mathbf{w}}
\newcommand{\bfzero}{\mathbf{0}}
\newcommand{\bfnabla}{{\boldsymbol\nabla}}
\newcommand{\Laplace}{\bfnabla^2}
\newcommand{\tr}{\operatorname{tr}}
\newcommand{\Ccc}{\operatorname{Ccc}}
\newcommand{\calD}{\mathcal{D}}
\newcommand{\barA}{{\bar{A}}}
\newcommand{\barC}{{\bar{C}}}
\newcommand{\barP}{{\bar{P}}}
\newcommand{\barT}{{\bar{T}}}
\newcommand{\barmu}{{\bar{\mu}}}
\newcommand{\barlambda}{{\bar{\lambda}}}
\newcommand{\tilD}{{\tilde{D}}}
\newcommand{\tilG}{{\tilde{G}}}
\newcommand{\tilu}{{\tilde{u}}}
\newcommand{\hatQ}{{\hat{Q}}}

\begin{document}
%%% RCS formatting: do not change
\pagestyle{fancy}
\thispagestyle{plain}
\fancypagestyle{plain}{

%%%HEADER%%%
\fancyhead[C]{\includegraphics[width=18.5cm]{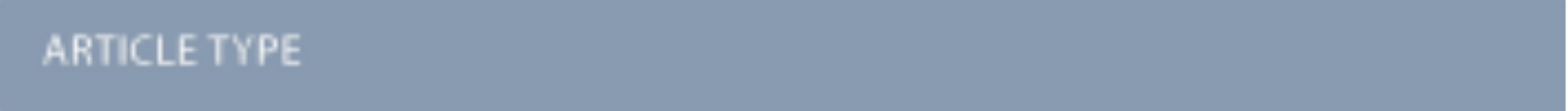}}
\fancyhead[L]{\hspace{0cm}\vspace{1.5cm}\includegraphics[height=30pt]{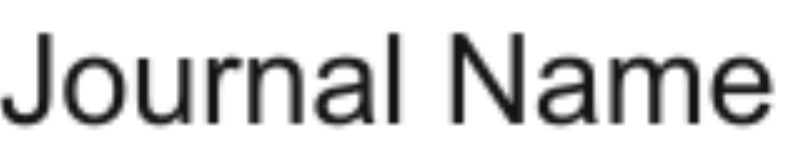}}
\fancyhead[R]{\hspace{0cm}\vspace{1.7cm}\includegraphics[height=55pt]{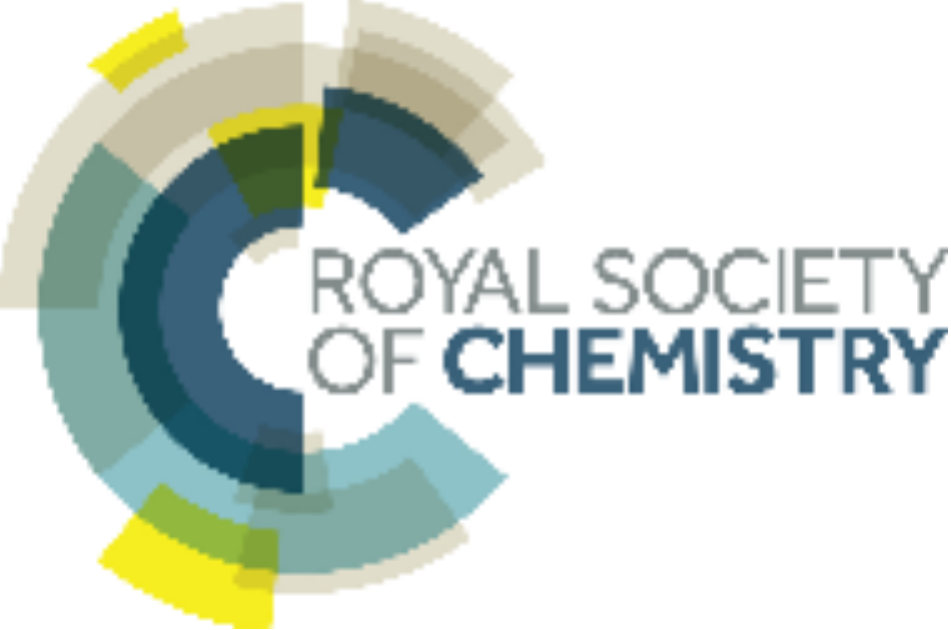}}
\renewcommand{\headrulewidth}{0pt}
}
%%%END OF HEADER%%%

%%%PAGE SETUP - Please do not change any commands within this section%%%
\makeFNbottom
\makeatletter
\renewcommand\LARGE{\@setfontsize\LARGE{15pt}{17}}
\renewcommand\Large{\@setfontsize\Large{12pt}{14}}
\renewcommand\large{\@setfontsize\large{10pt}{12}}
\renewcommand\footnotesize{\@setfontsize\footnotesize{7pt}{10}}
\makeatother

\renewcommand{\thefootnote}{\fnsymbol{footnote}}
\renewcommand\footnoterule{\vspace*{1pt}% 
\color{cream}\hrule width 3.5in height 0.4pt \color{black}\vspace*{5pt}} 
\setcounter{secnumdepth}{5}

\makeatletter
\renewcommand\@biblabel[1]{#1}
\renewcommand\@makefntext[1]%
{\noindent\makebox[0pt][r]{\@thefnmark\,}#1}
\makeatother 
\renewcommand{\figurename}{\small{Fig.}~}
\sectionfont{\sffamily\Large}
\subsectionfont{\normalsize}
\subsubsectionfont{\bf}
\setstretch{1.125} %In particular, please do not alter this line.
\setlength{\skip\footins}{0.8cm}
\setlength{\footnotesep}{0.25cm}
\setlength{\jot}{10pt}
\titlespacing*{\section}{0pt}{4pt}{4pt}
\titlespacing*{\subsection}{0pt}{15pt}{1pt}
%%%END OF PAGE SETUP%%%

%%%FOOTER%%%
\fancyfoot{}
\fancyfoot[LO,RE]{\vspace{-7.1pt}\includegraphics[height=9pt]{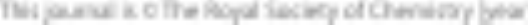}}
\fancyfoot[CO]{\vspace{-7.1pt}\hspace{13.2cm}\includegraphics{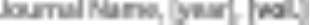}}
\fancyfoot[CE]{\vspace{-7.2pt}\hspace{-14.2cm}\includegraphics{head_foot/RF}}
\fancyfoot[RO]{\footnotesize{\sffamily{1--\pageref{LastPage} ~\textbar  \hspace{2pt}\thepage}}}
\fancyfoot[LE]{\footnotesize{\sffamily{\thepage~\textbar\hspace{3.45cm} 1--\pageref{LastPage}}}}
\fancyhead{}
\renewcommand{\headrulewidth}{0pt} 
\renewcommand{\footrulewidth}{0pt}
\setlength{\arrayrulewidth}{1pt}
\setlength{\columnsep}{6.5mm}
\setlength\bibsep{1pt}
%%%END OF FOOTER%%%

%%%FIGURE SETUP - please do not change any commands within this section%%%
\makeatletter 
\newlength{\figrulesep} 
\setlength{\figrulesep}{0.5\textfloatsep} 

\newcommand{\topfigrule}{\vspace*{-1pt}% 
\noindent{\color{cream}\rule[-\figrulesep]{\columnwidth}{1.5pt}} }

\newcommand{\botfigrule}{\vspace*{-2pt}% 
\noindent{\color{cream}\rule[\figrulesep]{\columnwidth}{1.5pt}} }

\newcommand{\dblfigrule}{\vspace*{-1pt}% 
\noindent{\color{cream}\rule[-\figrulesep]{\textwidth}{1.5pt}} }
\makeatother
%%%END OF FIGURE SETUP%%%

% terrible RCS "maketitle": <<<
\twocolumn[
  \begin{@twocolumnfalse}
\vspace{3cm}
\sffamily
\begin{tabular}{m{4.5cm} p{13.5cm} }

\includegraphics{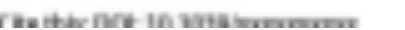} & \noindent\LARGE{\textbf{The range and nature of effective interactions in hard-sphere solids}} \\
\vspace{0.3cm} & \vspace{0.3cm} \\

 & \noindent\large{Michael Schindler$^{\ast}$\textit{$^{a}$} and A.\,C.\ Maggs\textit{$^{a}$}} \\

\includegraphics{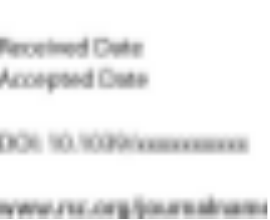} & \noindent\normalsize{%
  Colloidal systems observed in video microscopy are often analysed using the
  displacements correlation matrix of particle positions. In non-thermal
  systems, the inverse of this matrix can be interpreted as a pair-interaction
  potential between particles. If the system is thermally agitated, however,
  only an effective interaction is accessible from the correlation matrix. We
  show how this effective interaction differs from the non-thermal case by
  comparing with high-statistics numerical data from hard-sphere crystals.} \\

\end{tabular}

\end{@twocolumnfalse} \vspace{0.6cm}
]

\renewcommand*\rmdefault{bch}\normalfont\upshape
\rmfamily

\footnotetext{\textit{$^{a}$~UMR Gulliver 7083 CNRS, ESPCI ParisTech, PSL Research University, 10~rue Vauquelin, 75005~Paris, France}}
%\footnotetext{\textit{$^{b}$~Address, Address, Town, Country. }}
%%% >>>

\section{Introduction}
% <<<
% What is done: correlation matrix, inversion
% very much liked in studies of jammed or glassy dynamics of soft and hard particles
The analysis of the displacement--displacement correlation function between
many particles has become a standard tool used on experimental
data~\cite{ghosh,antinaPRL,pen,arjun,hard} acquired by video and confocal microscopy as well as on numerical
data.\cite{brito,silke,andrea,claire} It is much, but not exclusively, used
by researchers interested in jammed and glassy dynamics of soft particles. In
practice, one first determines the reference positions of particles and then
records the correlations of displacements with respect to these reference
positions for all particles. The averages are done either over time, using
particle trajectories, or over an ensemble, using Monte-Carlo sampling. The
result is a large correlation matrix~$G$ between all displacements of all
particles, which is then inverted to give another matrix~$D$.

% dynamical matrix and hessian
This matrix~$D$ is often interpreted as an effective interaction matrix which
characterises oscillations of the particles around their reference positions.
In that context, $D$~is called the \emph{dynamical matrix}.\cite{Born98,silke}
This interpretation has its origins in the analysis of non-thermal systems
where massive particles are connected by harmonic or other interaction
potentials. The dynamical matrix is then proportional to the Hessian matrix~$H$
of second derivatives of the total interaction potential\cite{silke} (with the
particle mass and with $kT$ as proportionality factors, which will play no role
in the following). When activated by a small external stimulus, the response of
the system is a mixture of oscillations which are given by the eigenfrequencies
and eigenvectors of the dynamical matrix. One condition for this analysis to
remain true is that the oscillation amplitudes are small, so that the
dynamical matrix does not depend on the displacements. With the masses of the
particles and the measured correlation matrix~$G$ as input, one can obtain information about the
interaction potential between the particles.

% dynamical matrix: from athermal to thermally decorated systems.
The protocol to record data and to invert the matrix~$G$ works equally well on
thermal as on non-thermal systems. In contrast to the Hessian matrix~$H$, which
contains the static potential interactions, the matrix~$G^{-1}$ contains also
all the thermal contributions, including entropic. How is the dynamical matrix
changed by these contributions? To what extent can we still infer the
interaction between particles? For example, if the true interaction acts only
between nearest neighbours, the same is true for~$H$, but does it also hold for
the effective interactions obtained from~$G^{-1}$?

% intermediate levels
% TODO make this a footnote?
Between these two limits of completely non-thermal~$H$ and fully
thermal~$G^{-1}$, a plethora of intermediate levels can be identified, which
account for more or fewer aspects of the thermal distribution and which may
serve for more or less reasonable models for the observed system. We refer to
\citet{silke} for a review of some of these intermediate levels and the
connections between them, such as the \emph{harmonic approximation} or the
\emph{shadow system}. As we mention \emph{models} for the system, it is
important to note that the Hessian matrix with which we compare $G^{-1}$ may
derive from modelling potentials very different from the true ones.
% Nomenclature HenBriDau12:
%  "empirical correlation matrix"
%  "true correlation matrix"
%  "Hessian matrix of pair-potentials" -- in terms of {r}
%     ... but then used in terms of {u}
%  "stiffness matrix" == "Hessian matrix"
%  "dynamical matrix": Hessian / mass
%
% *  The shadow system reproduces the same correlation matrix C as the true
%    system (by construction).
%
%    The "shadow system" is often thought of as a system with
%    particles connected by springs, the spring constants contained in the
%    "effective stiffness matrix" C^{-1}.
%
%    It is formulated as if connecting every particle with every other particle
%    were enough to correctly reproduce the correlation matrix C.
%    This is in fact wrong. When C^{-1} is used, the interaction between each
%    pair of particles is a dxd matrix and can thus contain nonparallel components.
%    (This is a problem of words only, not of the equations)

% the present paper
The aim of the present paper is to establish the difference between the two
extreme cases, which are the non-thermal case (Hessian matrix~$H$) and the
fully thermal matrix~$G^{-1}$ obtained from observation data. We here want to
see the largest difference and therefore choose as system of observation a
crystal of hard spheres which collide elastically. It is usually said that
thermal (\emph{entropic}) effects are maximal in the hard-sphere system. We
also have to choose a model system of which we calculate the Hessian matrix. We
here take the most general form of pairwise, central potential interactions.

% elastic moduli
Beside its interpretation as effective interactions, the matrix~$G^{-1}$ -- and
more directly the correlation matrix~$G$ -- serve also to extract elastic
moduli from thermal fluctuations. Because the elastic moduli are macroscopic
quantities of the whole system, this measurement involves only large-wavelength
properties of the effective interactions contained in~$G^{-1}$. Of course, the
same procedure can be done in the non-thermal case using $H$ instead of
$G^{-1}$. One advantage of the elastic moduli is that their calculation is not
limited to smooth pairwise potentials but applies to other types of interaction
as well, such as strictly hard spheres. We expect to see here another
discrepancy between the thermal and the non-thermal cases: When one searches
the long-wavelength limit of the Hessian matrix, one obtains the so-called
\emph{Cauchy relations} which reduce the number of independent elastic
constants in materials below the number expected from symmetry considerations.
For example, in two-dimensional hexagonal crystals and in three-dimensional
cubic crystals the Cauchy relations can be written as an equality $\lambda=\mu$
between two Lamé coefficients. The Cauchy relations apply if interactions
between atoms (or colloids) are pairwise, central, sufficiently short ranged,
and if thermal fluctuations are neglected. The role of thermal fluctuations was
clarified by \citet{hoover} who calculated the extra contributions (beyond
those found by \citet{Born98}) in the expressions for the elastic constants
at nonzero temperature. These contributions can be seen as the ``entropic''
part of the elastic moduli, and they explain why hard-sphere crystals do not
satisfy the Cauchy relations~\cite{WojTreBraKow03}.

The paper is structured as follows: In section~\ref{sec:theory} we collect some
results from known theories, in particular from the non-thermal dynamical
(Hessian) matrix. We derive some properties of the dynamical matrix which allow
comparison with the fully thermal case. The numerical data for the hard-sphere
crystals is then presented in section~\ref{sec:numerics} in one, two and three
dimensions. Finally, in section~\ref{sec:discussion} we discuss the observed
differences between the two cases.
% >>>

\section{Effective interactions and the central force model}
\label{sec:theory}
% <<<
The simplest (Cauchy) model of elasticity in the physics of solids supposes
that interactions are pairwise and central between all particles in the solid,
deriving from a scalar potential, see for example \textsection~29 of the book
by~\citet{Born98}. In this section we study the elastic properties of such
systems, in order to better understand where the numerical data demonstrate the
presence of non-central potentials.

In this model, the total potential energy~$\Phi$ of a configuration is a
function of all particle positions~$\bfr_\bfi$. It is expressed as a sum of
central pair-potentials $g_{\bfi\bfj}$,
\begin{equation}
  \label{model:Phi}
  \Phi(\{\bfr\}) = \sum_{(\bfi\bfj)} g_{\bfi\bfj}\bigl((\bfr_\bfi-\bfr_\bfj)^2/2\bigr)
\end{equation}
The sum runs over all different pairs of particles. To simplify the algebra and
avoid square roots in the calculations, the $g_{\bfi\bfj}$~are functions of the
distance squared. The functions should be identical for symmetry-related pairs
of particles, but otherwise each function is independent.

In order to unify the notation with the numerical simulations of later
sections, we assume already here that the particles are in a crystalline
configuration (hexagonal or face-centered cubic, depending on the number of
spatial dimensions). Throughout the paper, $\bfi,\bfj,\bfm,\bfn$ denote tuples
of integer indices on the Bravais lattice and serve to identify the sites on
the lattice. Spatial indices are denoted by Greek letters. The spatial
(Euclidean) positions of lattice sites are denoted by vectors $\bfR_\bfi$. They
are multiples of the lattice spacing~$d_0$. The numerical, and optionally also
the theoretical calculations, are periodic in space. All differences between
lattice indices, $\bfi-\bfj$, or lattice positions,
$\bfR_{\bfi\bfj}:=\bfR_{\bfi-\bfj}=\bfR_\bfi-\bfR_\bfj$ are thus understood
\emph{modulo the periodicity}, the result being mapped back into a centered
copy of the periodic box.
% >>>
\subsection{Description in terms of local displacements}% <<<
\label{sec:local}

We now calculate the Hessian matrix of the potential energy~$\Phi(\{\bfr\})$
around the crystalline reference state. This amounts to small-amplitude
displacements, which are non-thermal in their nature. The reference state has
inversion symmetry (hexagonal in 2D, face-centered cubic in 3D), which imposes
that the partial first derivatives vanish when evaluated at the reference
positions:
\begin{equation}
  \label{vanish}
  \frac{\partial\Phi}{\partial r_{\bfi\alpha}}\bigl(\{\bfR\}\bigr) = 0
\end{equation}
for all $\bfi$ and $\alpha$. From Eq.~\eqref{vanish} it follows that the
increase in potential energy of the system due to local displacements is given
to quadratic order by
\begin{equation}
  \label{DeltaPhi}
  \Delta\Phi := \Phi\bigl(\{\bfr\}\bigr) - \Phi\bigl(\{\bfR\}\bigr)
    = \frac{1}{2} \sum_{\alpha,\beta,\bfi,\bfj} H_{\alpha\beta}(\bfj-\bfi)\, u_{\bfi\alpha}\, u_{\bfj\beta}
\end{equation}
where $\bfu_\bfi=\bfr_\bfi-\bfR_\bfi$ is the displacement vector of
particle~$\bfi$. The Hessian matrix consists of second
derivatives, evaluated at the reference positions,
\begin{equation}
  H_{\alpha\beta}(\bfj-\bfi) := \frac{\partial^2\Phi}{\partial r_{\bfi\alpha} \partial r_{\bfj\beta}}\bigl(\{\bfR\}\bigr).
\end{equation}
Start with the first derivative
\begin{equation}
  \frac{\partial\Phi}{\partial x_\bfi}\bigl(\{\bfr\}\bigr) = \sum_{\bfj\neq\bfi} (x_\bfi -x_\bfj) g'_{\bfi\bfj}
\end{equation}
and obtain $H_{\alpha\beta}(\bfj-\bfi)$ for $\bfi\neq\bfj$:
\begin{equation}
  H_{\alpha\beta}(\bfj-\bfi) = -(R_{\bfi\alpha}{-}R_{\bfj\alpha})(R_{\bfi\beta}{-}R_{\bfj\beta})\,g''_{\bfi\bfj} - \delta_{\alpha\beta}\, g'_{\bfi\bfj},
\end{equation}
We abbreviate our notation by introducing
$\bfR_{\bfj-\bfi}=\bfR_\bfj-\bfR_\bfi$ and $g_{\bfj-\bfi}=g_{\bfi\bfj}$ and
obtain
\begin{subequations}
\begin{equation}
  H_{\alpha\beta}(\bfn) = -R_{\bfn\alpha} R_{\bfn\beta}\,g''_{\bfn}\bigl(R_\bfn^2/2\bigr) - \delta_{\alpha\beta}\, g'_{\bfn}\bigl(R_\bfn^2/2\bigr).
\end{equation}
At $\bfj=\bfi$, the matrix is constrained by the fact that the sum over all $\bfn$
vanishes,
\begin{equation}
  \label{sumconstraint}
  H_{\alpha\beta}(\bfzero) = -\sum_{\bfn\neq\bfzero} H_{\alpha\beta}(\bfn).
\end{equation}
\end{subequations}

Let us return for a moment to the meaning of this Hessian matrix. As was
mentioned in the introduction, it is also the dynamical matrix of particles (of
unit mass), in the sense that their displacements oscillate around the
reference state according to the equation
\begin{equation}
  \ddot\bfu_{\bfi\alpha} = -\sum_{\bfj,\beta}H_{\alpha\beta}(\bfj-\bfi) \bfu_{\bfj\beta}.
\end{equation}
In words, the \emph{interaction matrix} $H_{\alpha\beta}(\bfi-\bfj)$ governs
the force exerted on particle~$\bfi$ by particle~$\bfj$. For every separation
vector $\bfn$ between two particles, $H_{\alpha\beta}(\bfn)$ is a
$d{\times}d$-matrix. We will find it important to analyse the forces and
displacements parallel and orthogonal to the separation vector~$\bfn$. In
particular in two dimensions, we can use the rotated orthonormal basis
$(\hat\bfR_\bfn,\hat\bfo)$, where the hat denotes a normalised vector, to
define the rotated components of the matrix~$H(\bfn)$. We find
\begin{subequations}
\label{Hrot}
\begin{align}
  \label{Hrot:parall}
  H_\parallel(\bfn)
    &:= \hat\bfR_\bfn^T H(\bfn)\hat\bfR_\bfn
      = -\bfR_\bfn^2 g''_{\bfn}(\bfR_\bfn^2/2)  - g'_{\bfn}(\bfR_\bfn^2/2) \\
  \label{Hrot:ortho}
  H_\perp(\bfn)
    &:= \hat\bfo^T H(\bfn)\hat\bfo
      = - g'_{\bfn}(\bfR_\bfn^2/2) \\
  \label{Hrot:sym}
  H_\text{sym} (\bfn)
    &:= \bigl(\hat\bfR_\bfn^T H(\bfn)\hat\bfo + \hat\bfo^T H(\bfn)\hat\bfR_\bfn\bigr) / 2
      = 0 \\
  \label{Hrot:asym}
  H_\text{asym}(\bfn)
    &:= \bigl(\hat\bfR_\bfn^T H(\bfn)\hat\bfo - \hat\bfo^T H(\bfn)\hat\bfR_\bfn\bigr) / 2
      = 0
\end{align}
\end{subequations}%
The components~$H_\parallel$ and $H_\perp$ control those components of the
force which are parallel to the displacements, respectively both parallel
to~$\bfn$ ($H_\parallel$) or both orthogonal to it ($H_\perp$). The
(a)symmetric parts govern those components of the force which are orthogonal to
the displacements they result from. They are found to vanish in the present
model, meaning that displacements can generate only forces which are parallel
to them. This property appears as a result of the potentials being central, and
it will be tested on the numerical data in the sections below.

We can also examine the results in the non-rotated frame. In particular, for
the off-diagonal terms we find
\begin{equation}
  \label{H:xy}
  H_{xy}(\bfn) = -X_\bfn Y_\bfn g'_\bfn\bigl(R_\bfn^2/2\bigr).
\end{equation}
which is an odd function with respect to $X_\bfn$ and to $Y_\bfn$. Also this
symmetry will be subject to comparison with the numerical data.
% >>>
\subsection{Long-wavelength limit}% <<<

The elastic tensor for the model~\eqref{model:Phi} has been calculated
by~\citet{Born98} (\textsection~29). In our notation, their result for the
elastic tensor reads
\begin{equation}
  \label{barC}
  \barC_{\alpha\sigma\beta\tau}
    = \frac{N}{2V}\sum_{\bfn} R_{\bfn,\alpha} R_{\bfn,\sigma} R_{\bfn,\beta} R_{\bfn,\tau}\,
      g^{\prime\prime}_{\bfn}\bigl(R_\bfn^2/2\bigr).
\end{equation}
Here, $N$ is the number of particles, and $V$~is the volume of the (periodic)
box. We use an overbar to distinguish this tensor from the true (thermal)
elastic tensor. Expression~\eqref{barC} is symmetric under all index
permutations, which is higher than the symmetry required for the elastic tensor.
This observation leads us directly to the Cauchy relation: The general form of
the elastic tensor is given by group theory, for example in a two-dimensional
hexagonal lattice,
\begin{equation}
  \label{C}
  C_{\alpha\beta\sigma\tau} = \lambda\delta_{\alpha\beta}\delta_{\sigma\tau}
  + \mu\bigl(\delta_{\alpha\sigma}\delta_{\beta\tau}
           + \delta_{\alpha\tau}\delta_{\beta\sigma}\bigr),
\end{equation}
with the Lamé coefficients $\lambda$ and~$\mu$. This expression becomes
completely symmetric under index permutations only if the Cauchy
relation~$\lambda=\mu$ is satisfied. The situation is similar in
three-dimensional cubic lattices, only that there is another term in
Eq.~\eqref{C} with the anisotropy coefficient as prefactor. There, the Cauchy
relation reduces three independent coefficients to two. In
Section~\ref{sec:numerics} we will measure the Lamé coefficients from numerical
data and test directly whether the Cauchy relation is satisfied.

When we determine the Lamé coefficients from Eq.~\eqref{barC}, we find indeed
that they are identical,
\begin{align}
  \notag
  \barlambda = \barmu
   &= \frac{N}{16V} \sum_{\bfn} R_\bfn^4 g^{\prime\prime}_{\bfn} \\
  \label{barmu}
   &= \frac{N}{16V} \sum_{\bfn} R_\bfn^2 \bigl(H_\perp(\bfn)-H_\parallel(\bfn)\bigr).
\end{align}
In the last step we made use of Eqs.~\eqref{Hrot:parall} and \eqref{Hrot:ortho}.

We must mention that the calculation by \citet{Born98} is done with a
stress-free reference state in mind. In model~\eqref{model:Phi} we do not know
\emph{a priori} whether the stress vanishes in the reference state or not, and
for the hard-sphere crystals in the sections below, we know that they stay
crystalline only if they are sufficiently compressed by the periodic box. We
therefore need to consider the general case of a non-vanishing stress in the
reference state. The effect of this stress is that the elastic tensor~$C$,
which characterises the quadratic increase of free energy with strain, is not
identical to the elastic tensor~$A$ which characterises the motion of
long-wavelength vibrations and thermal excitations.\cite{Wallace70} The two
tensors are related by
\begin{gather}
  \label{AandC}
  A_{\alpha\beta\sigma\tau} := C_{\alpha\beta\sigma\tau} + T_{\beta\tau}\delta_{\alpha\sigma},
\end{gather}
where $T$~denotes the stress tensor in the reference state. To obtain the
stress tensor, we start with the force between two particles,
\begin{equation}
  \bff_{\bfi\bfj} = - (\bfR_\bfi-\bfR_\bfj) g'_{\bfi\bfj}.
\end{equation}
The virial part of the stress tensor is thus
\begin{align}
  \notag
  \barT_{\sigma\tau}
    = -\frac{1}{2V}&\sum_{(\bfi\bfj)} {(\bfR_\bfi-\bfR_\bfj)}_\sigma f_{\bfi\bfj,\tau} \\
    \label{barT}
    = \frac{N}{2V}&\sum_{\bfn} R_{\bfn,\sigma} R_{\bfn,\tau} g'_\bfn.
\end{align}
This expression has been given also by~\citet{Born98}.

It is desirable to derive expression~\eqref{barC} directly from a
long-wavelength limit of the interaction matrix~$H$. This calculation cannot be
done at this point, since we require thermal fluctuations which then yield the
tensor~$A$. From there we can conclude on the tensor~$C$. We therefore defer
this calculation to section~\ref{sec:hexcontinuum} below, where we will do it
on the matrix~$D$. When we replace $D(\bfn)$ by $H(\bfn)/kT$ in the
result~\eqref{limitD} there, we find the equivalent of the elastic tensor~$A$,
\begin{multline}
  \label{barA}
  \barA_{\alpha\sigma\beta\tau}
   := -\frac{N}{2V}\sum_{\bfn} R_{\bfn,\sigma} R_{\bfn,\tau} H_{\alpha\beta}(\bfn) \\
    = \frac{N}{2V}\sum_{\bfn} R_{\bfn,\sigma} R_{\bfn,\tau}
      \bigl(R_{\bfn,\alpha} R_{\bfn,\beta}\,g^{\prime\prime}_{\bfn} + \delta_{\alpha\beta}\,g^\prime_{\bfn}\bigr),
\end{multline}
which is entirely compatible with Eqs.~\eqref{barC},
\eqref{AandC} and~\eqref{barT}.
% >>>
\subsection{Description in terms of global deformations}% <<<
\label{sec:hoover}%
In the above section we expressed the elastic constants in terms of a
long-wavelength limit of the matrix~$H$. This matrix, as it involves second
derivatives of the potential energy only, is non-thermal in nature. At least
for the elastic constants, it is possible to include thermal
effects\cite{hoover}. We introduce a parameter for global deformations of the
system, the strain~$\eta$ and perform a Taylor expansion of the Free Energy
$F(\eta)=-kT\ln(Z(\eta))$ about zero strain ($\eta=0$). As the whole box is
changed parametrically, we could speak of a ``zero-wavevector'' ($|\bfQ|=0$)
method instead of the long-wavelength limit $|\bfQ|\to0$. The resulting stress
tensor and elastic tensor are expressed by the usual Gibbs-weighted average
over configurations,\cite{hoover}
\begin{align}
  T_{\alpha\beta}
   \notag
   &= \frac{1}{V}\frac{\partial F}{\partial\eta_{\alpha\beta}}(0) \\
   \label{hoover:stress}
   &= -kT\frac{N}{V} \delta_{\alpha\beta}
    + \frac{1}{V} \Bigl\langle {\textstyle\sum\limits_{(\bfi\bfj)}}\: g_{\bfi\bfj}^{\prime}\bigl(r_{\bfi\bfj}^2/2\bigr)\, r_{\bfi\bfj,\alpha}\,r_{\bfi\bfj,\beta}\Bigr\rangle \\
  C_{\alpha\beta\sigma\tau}
   \notag
   &= \frac{1}{V}\frac{\partial^2 F}{\partial\eta_{\alpha\beta}\partial\eta_{\sigma\tau}}(0)
    = 2kT \frac{N}{V} \delta_{\alpha\tau}\delta_{\beta\sigma} \\
   \notag
   &- \frac{1}{V\,kT} \Ccc\Bigl({\textstyle\sum\limits_{(\bfi\bfj)}}\: g_{\bfi\bfj}^{\prime}\bigl(r_{\bfi\bfj}^2/2\bigr)\, r_{\bfi\bfj,\alpha}\,r_{\bfi\bfj,\beta},\;
                                              {\textstyle\sum\limits_{(\bfi\bfj)}}\: g_{\bfi\bfj}^{\prime}\bigl(r_{\bfi\bfj}^2/2\bigr)\, r_{\bfi\bfj,\sigma}\,r_{\bfi\bfj,\tau} \Bigr) \\
   \label{hoover:elast}
   &+ \frac{1}{V} \Bigl\langle {\textstyle\sum\limits_{(\bfi\bfj)}}\: g_{\bfi\bfj}^{\prime\prime}\bigl(r_{\bfi\bfj}^2/2\bigr)\, r_{\bfi\bfj,\alpha}\,r_{\bfi\bfj,\beta}\,r_{\bfi\bfj,\sigma}\,r_{\bfi\bfj,\tau}\Bigr\rangle
\end{align}
The function $\Ccc(a,b):=\langle ab\rangle - \langle a\rangle\langle b\rangle$
is a cumulant-like cross-correlation function. The terms on the right-hand side
of Eq.~\eqref{hoover:stress} are called the \emph{kinetic} and the
\emph{virial} terms. The terms in Eq.~\eqref{hoover:elast} are called the
\emph{kinetic term}, the \emph{fluctuations term}, and the \emph{Born term}.\cite{hoover}

The virial term in~\eqref{hoover:stress} and the Born term
in~\eqref{hoover:elast} resemble much the expressions we found for
stress~\eqref{barT} and elastic tensor~\eqref{barC} from
model~\eqref{model:Phi}. Only, there we evaluated at the reference positions,
and here we average over positions around these reference positions. Of course,
if the position distribution is very narrow, the values obtained from both can
be close. Both kinetic terms and the fluctuation term are missing, however. The
absence of these terms is where the tensor~$\barC$ has its higher symmetry
from, and why we obtained the Cauchy relation for it.

The elastic tensor~\eqref{hoover:elast} includes all thermal effects, because
we take the second derivative of the free energy. A similar derivative of the
potential energy~\eqref{model:Phi} would simply yield again the Born-like term
in~\eqref{barC}. It would be very nice to add thermal effects to the
interaction matrix, just by replacing the potential energy by the free energy
in the second derivative with respect to the reference positions~$\bfR_\bfi$.
Unfortunately, the reference positions are not parameters, they appear by a
spontaneous symmetry breaking, and we therefore do not know how to do such a
derivative. For the moment, we can only work with the imperfect expressions of
the above sections and compare them with full numerical data.
% >>>

\section{Numerical results from Hard sphere simulations}
\label{sec:numerics}
% <<<
We performed \emph{high statistics} event-driven molecular dynamics simulations
of $N$~hard spheres in one, two and three dimensions ($d=1,2,3$). In one
dimension the particles were confined to a line. In two dimensions we simulated
a hexagonal crystal within a box adapted to the Bravais lattice. In three
dimensions we performed simulations of face-centred cubic crystals, again
within a box adapted to the Bravais lattice. In all dimensions the simulation
box was continued periodically. During a simulation the total momentum and the
(kinetic) energy were conserved. The densities were always so high that no
particle interchange occurred over the simulation time, so that particle
diffusion and defects can be neglected in the data analysis. The Bravais
lattice thus defines the reference positions of the spheres, and these coincide
with the average positions.

For all numerical data the units are chosen such that the particles have unit
diameter, unit mass, and that $kT=1$ with $T$~the temperature.

During the simulation we took a series of $K$~data recordings. For each
recording, we measure the deviations $\bfu_\bfi:=\bfr_\bfi-\bfR_\bfi$ of the
instantaneous particle positions from their reference positions and calculate
the $dN{\times}dN$~correlation matrix
\begin{equation}
  G_{(\bfi,\alpha)(\bfj,\beta)} := \frac{1}{K} \sum_{k=1}^K u_{\bfi\alpha} u_{\bfj\beta}
\end{equation}
The matrix has a simple physical interpretation from linear response theory: It
determines the vectorial displacement at $\bfj$ due to a force at~$\bfi$. We
will thus call it a Green function in the following. In order to reduce
statistical noise, we average elements of~$G$ which are related by the
translational symmetry of the lattice,
\begin{equation}
  G_{\alpha\beta}(\bfn) := \frac{1}{N} \sum_{\bfi} G_{(\bfi,\alpha)(\bfi+\bfn,\beta)}.
\end{equation}
\begin{figure}%
  \centering
  \includegraphics{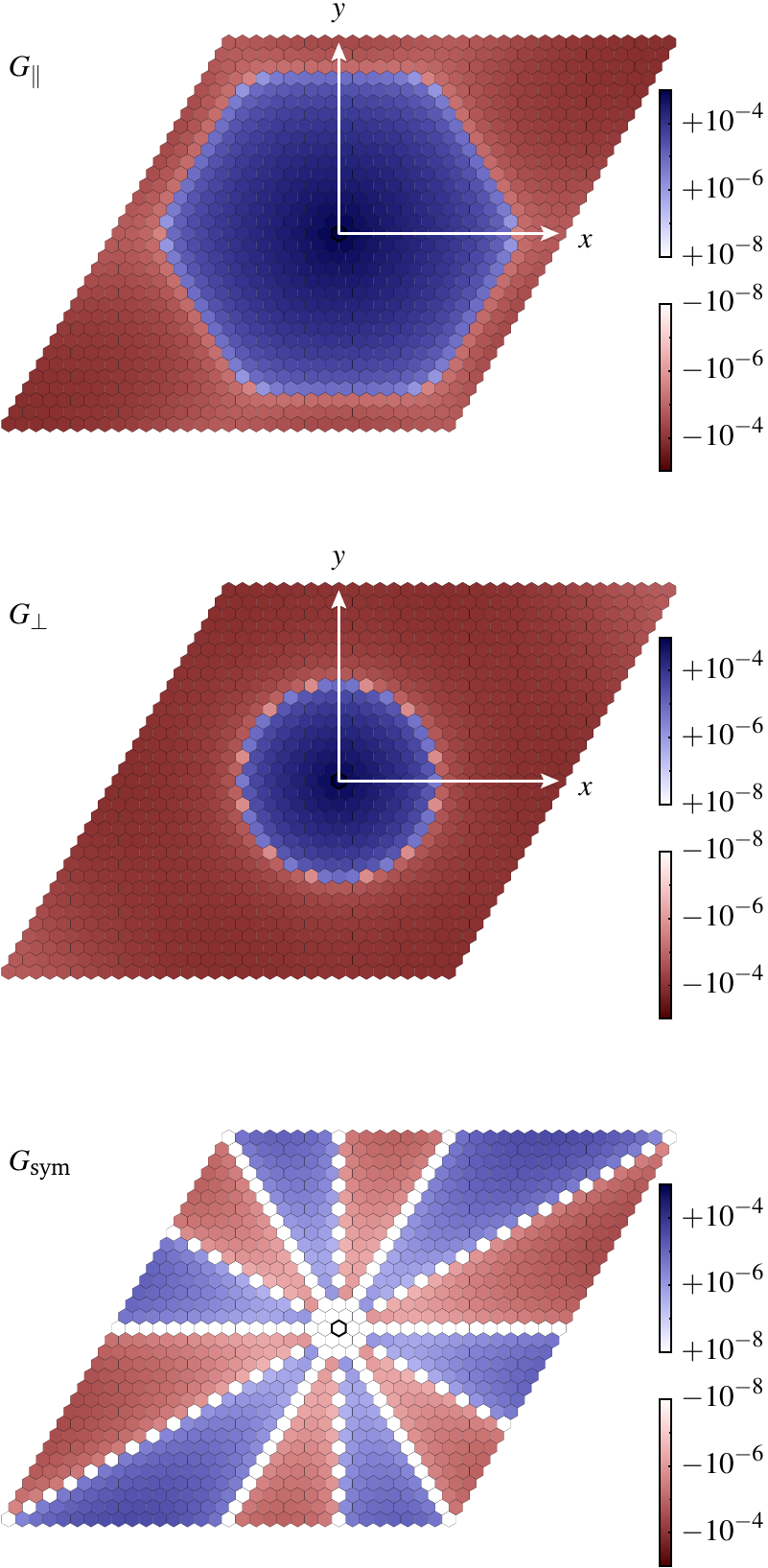}%
  \caption{The two-dimensional Green function in a periodic $32{\times}32$
  hexagonal lattice. Plotted are the components of the $2{\times}2$~matrices
  $G_{\alpha\beta}(\bfn)$ in a rotated orthonormal basis $(\hat\bfR_\bfn,
  \hat\bfo)$: $G_\parallel=\hat\bfR_\bfn^T G\hat\bfR_\bfn$, $G_\perp=\hat\bfo^T
  G\hat\bfo$, $G_\text{sym}=(\hat\bfR_\bfn^T G\hat\bfo + \hat\bfo^T
  G\hat\bfR_\bfn)/2$. The antisymmetric off-diagonal terms vanish to within statistical
  noise~($\pm10^{-9}$). In the lowest panel the white lines correspond to a
  mechanical response parallel to an imposed force.}%
  \label{fig:G}%
\end{figure}%
The sum runs over all sites of the Bravais lattice. We could also average over
the rotational and inversion symmetries, but we prefer to see the symmetries
appear from the data in order get a idea of the statistical errors; the
translational average appears to be sufficient for noise reduction. A visual
representation of the elements of $G(\bfn)$ in two dimensions is given in
Fig.~\ref{fig:G}. It is a rather structureless object, and it depends on the
size and shape of the periodic box due to the logarithmic nature of
two-dimensional Green functions. In the figure we plot the elements of
$G(\bfn)$ in a rotated coordinate system with one basis vector aligned with the
vector $\bfR_\bfn$. We note in particular the interesting physical structure
that occurs in the off-diagonal elements. These elements of the response
function vanish along high symmetry directions: a force along these symmetry
directions gives a purely parallel displacement. This gives rise to a set of
radial white lines in the lowest panel of Fig.~\ref{fig:G}.

We then use matrix algebra to numerically invert the large matrix~$G$ to
produce the effective interaction matrix~$D$. More precisely, we use the
translation-averaged $dN{\times}dN$ matrix
\begin{equation}
  G_{\alpha\beta}(\bfi,\bfj) := G_{\alpha\beta}(\bfj-\bfi).
\end{equation}
and take its Moore--Penrose pseudoinverse because we must avoid inversion of
the zero eigenvalues in~$G$ which come from momentum conservation. The inverse is
denoted by $D_{\alpha\beta}(\bfi, \bfj)$.
%Also in the inverse $dN{\times}dN$ matrix $D_{\alpha\beta}(\bfi, \bfj)$, we
%make use of translational symmetry to reduce sparsity and noise.
In the following, we will analyse the properties of the resulting $2{\times}2$
matrices
\begin{equation}
  D_{\alpha\beta}(\bfn) := \frac{1}{N} \sum_{\bfi} D_{\alpha\beta}(\bfi,\bfi+\bfn).
\end{equation}
%
% >>>
\subsection{One-dimensional system}% <<<
\begin{figure}%
  \centering
  \includegraphics{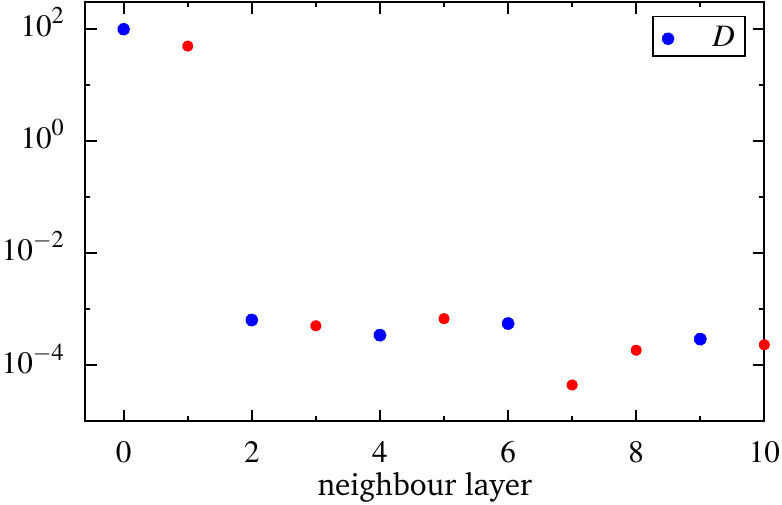}%
  \caption{The effective interactions $D(j-i)$ in a one-dimensional chain of
  $N=100$ impenetrable rods. Beyond nearest neighbours the interactions are
  zero to within statistical noise. Larger blue symbols are positive values,
  smaller red are symbols negative values. $K = N\times10^7$ recordings,
  obtained from three independently simulated trajectories.}%
\label{fig:1D}
\end{figure}

Previous work has led to a detailed understanding of the one-dimensional
problem.\cite{oned} We do not treat it in detail, but rather use it to check
some of the chain of data analysis. The result is very simple -- that the
effective interaction in one dimensional fluids is limited to nearest
neighbours; for larger separations $D$~is zero. We used our code for simulation
and data analysis and confirmed this results. Beyond the nearest neighbour, the
effective interaction $D(j-i)$ in Fig.~\ref{fig:1D} falls to zero within
statistical noise. Note that the relative statistical noise in this plot is at
the level of $10^{-5}$, showing the \emph{high precision} of the simulations.
Given the excellent results found in one dimension, we can feel confident that
the very different results found in two and in three dimensions are a result of
differing physics, and not problems related to systematic or statistical errors
in the data sets.

In the data of Fig.~\ref{fig:1D} we find the ratio between the two nonzero
values to be $D(0)/D(1) = -2.0\pm10^{-5}$. This is precisely the ratio that is
expected from a discretisation of the Laplace operator in terms of nearest
neighbours. We thus found the expected result, namely that the inverse of the
(static) Green function is the corresponding differential operator of
one-dimensional (static) elasticity.%
% >>>
\subsection{Two-dimensional crystal}% <<<
\begin{figure}%
  \centering
  \includegraphics{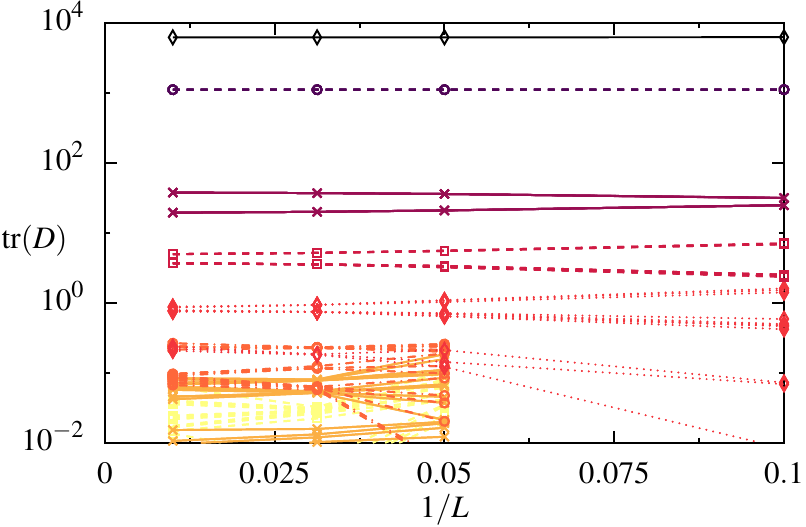}%
  \caption{Effective interactions as a function of the inverse system size
  $1/L$. Different plot styles correspond to different neighbour layers,
  starting from~$0$ at the top, down to~$7$. Fewer layers are plotted for
  $L=10$.}%
\label{fig:size}%
\end{figure}

We performed two-dimensional simulations in systems with $N=L{\times}L$ hard
disks, $L$~varying from $10$ to~$100$ with fixed surface fraction $\phi=0.85$.
We expect that the elements of the effective dynamical matrix~$D$ come from the
local physics (and available entropy) occurring near each particle and are not
the results of a non-trivial propagation of boundary effects down to the
microscopic scale. If the physics is local, we then expect that the values of
the elements of~$D$ vary very little with changes in the system size. To test
this, we plot in Fig.~\ref{fig:size} the evolution of $\tr\bigl(D(\bfn)\bigr)$
with the system size~$L$ and find that the lines for small separations~$\bfn$
are remarkably stable. This is true for all matrix elements of
$D_{\alpha\beta}$, not only for the trace. Since we are performing simulations
within a periodic box, it is clear that we should only be evaluating elements
for separations such that periodic copies do not contaminate the result. Thus
we will only quantitatively analyse~$D$ for separations which are smaller
than~$L/2$.

The correlation function~$G$, which is measured in the simulations and
displayed in Fig.~\ref{fig:G}, is subject to strong finite size corrections due
to the logarithmic nature of two-dimensional Green functions. The components of
the matrix~$D$, however, do not evolve for small values of~$1/L$. From the
curves in Fig.~\ref{fig:size} we decided to use a fixed value of $L=32$ for
further simulation in order to collect the highest possible statistics, while
being able to resolve effective interactions out to a large distance.%
\begin{figure}%
  \centering
  \includegraphics{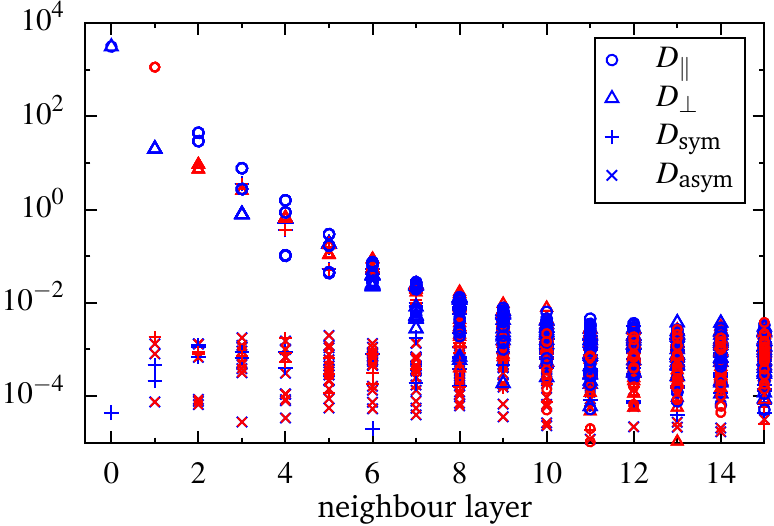}%
  \caption{Effective interactions in a hexagonal hard-disk crystal with $L=32$.
  Larger blue symbols are positive values, smaller red symbols are negative
  values. $K= 2N\times3.5{\times}10^6$ recordings, obtained from
  336~independent trajectories. Surface fraction~$\phi=0.85$. These data
  required around 300\,000 CPU core hours.}%
\label{fig:2D}
\end{figure}%

The resulting matrix elements are plotted in Fig.~\ref{fig:2D}. As we did
already in Fig.~\ref{fig:G}, we rotated the $2{\times}2$~matrices
$D_{\alpha\beta}(\bfn)$ for each separation vector~$\bfR_\bfn$. We used the
same rotated orthonormal basis and the same notation as we used in
Eqs.~\eqref{Hrot}. The interaction matrix depicted in Fig.~\ref{fig:2D} does
\emph{not} vanish beyond the first layer of neighbours -- very differently from
its unidimensional counterpart in Fig.~\ref{fig:1D}. We here observe
that~$D(\bfn)$ is not sparse. We used very high statistics in this plot to be
sure that the interaction data is well separated from statistical noise -- We
find it to be the case for the first six layers of neighbours. The noise level
can be read off from the values of $D_\text{asym}$ which are zero in noiseless
data, finding values below $2{\times}10^{-2}$. Beyond the 10th layer of
neighbours, signal and noise cannot be distinguished anymore. Another measure
of the statistical noise is the spread of those symbols which are equal by
symmetry of the lattice (smaller than the linewidth in the plot). Both measures
for the noise were observed to decrease while we accumulated more and more data
during the simulations.

We tried to characterise the decay of the effective interactions with particle
separation, which we can resolve out to the eighth layer of neighbours. We
tried fitting the data with both exponential and power law decays, plotting
both particle separation and layer number for the abscissa. No fit seemed
totally convincing with our data. If one insists on fitting with a power law
$\|\bfR_\bfn\|^{\alpha}$, one finds $\alpha\approx-6\pm0.5$. At the moment we
do not have analytic arguments that would predict the functional form of the
decay.
\begin{figure}%
  \centering
  \includegraphics{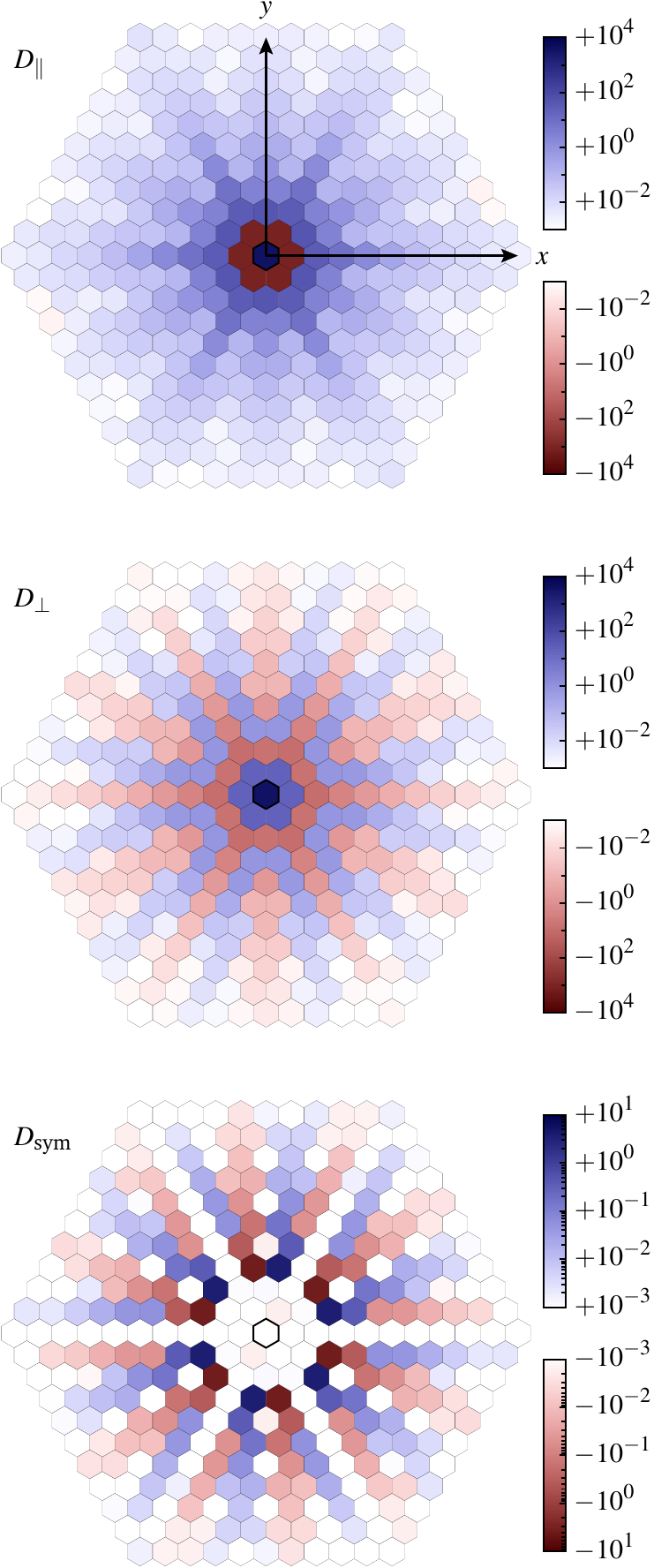}%
  \caption{Matrix elements of $D(\bfn)$ in rotated frames of reference, for
  several~$\bfR_\bfn$. The center of the hexagon corresponds to
  $\bfR_\bfn=\bfzero$. The antisymmetric off-diagonal $D_\text{asym}$ vanishes
  to within statistical noise. The reflection anti-symmetry of $D_{sym}$
  imposes that the elements of $D_{sym}$ are zero for the first two neighbour layers. All
  elements of $D_{sym}$ are small compared to $D_\parallel$ and $D_\perp $.}%
  \label{fig:dist_rot}%
\end{figure}%
\begin{figure}%
  \centering
  \includegraphics{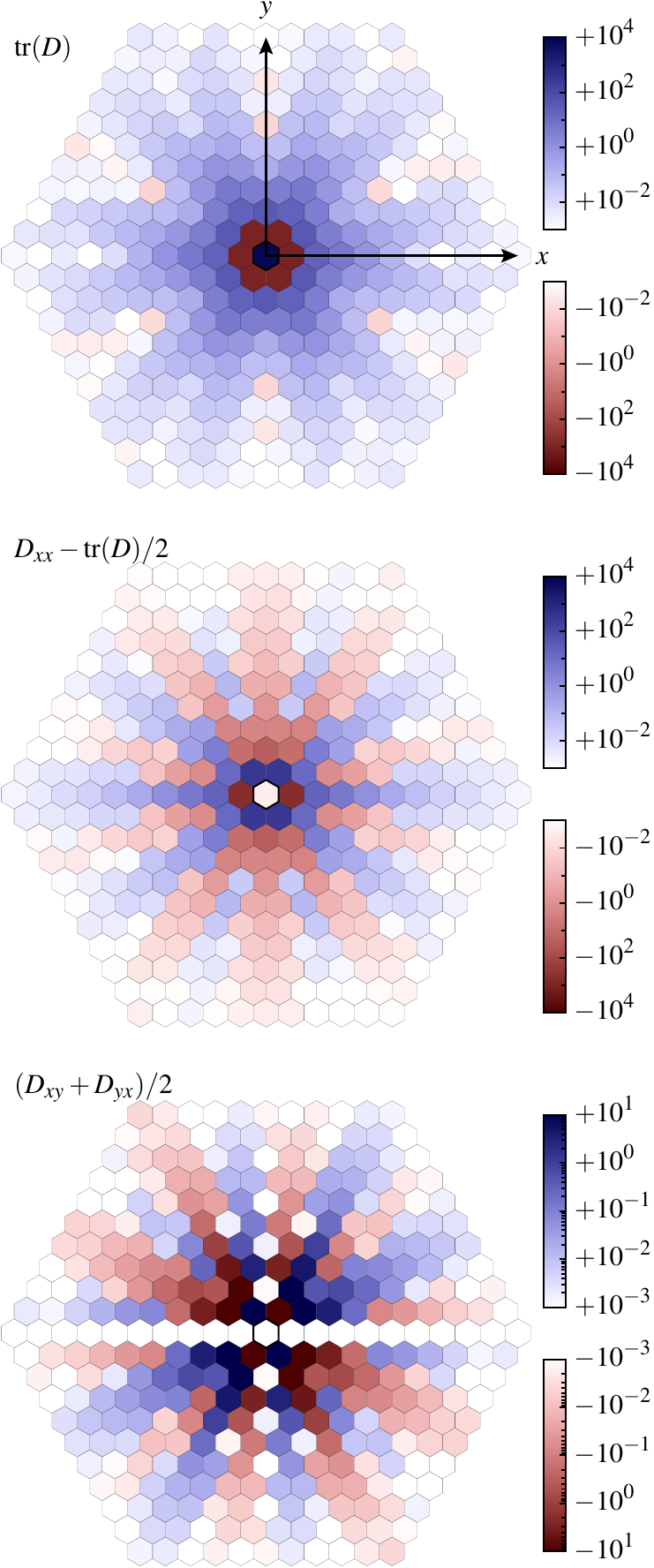}%
  \caption{Some linear combinations of matrix elements $D_{\alpha\beta}(\bfn)$
  for several~$\bfR_\bfn$. The antisymmetric off-diagonal $(D_{xy}-D_{yx})/2$
  vanishes to within statistical noise.\hfill\hbox to 5cm{} \hbox to 5cm{\vbox to 10.5mm{}}}
  \label{fig:norot}%
\end{figure}%
%

% >>>
\subsubsection{Two-dimensional visualisation}% <<<
We now give some more details of the data presented in Fig.~\ref{fig:2D}. In
particular, we like to visualise those points which are characterised by being
within same neighbour level, but which differ with respect to the hexagonal
symmetry.

We can study the $2{\times}2$~matrices $D(\bfn)$ in different frames of
reference. The first frame of reference is the orthonormal basis
$(\hat\bfR_\bfn,\hat\bfo)$ already used above. In Fig.~\ref{fig:dist_rot} we
plot the parallel, the orthogonal, and the off-diagonal component of the
matrix. All three panels are invariant under rotations of 60~degrees. Under
mirroring, however, only the first two are invariant while $D_\text{sym}$
changes sign. We did not plot $D_\text{asym}$ because it contains only noise.

A second manner of examining the data is in the fixed Cartesian frame, which is
the same for all~$\bfn$. In Fig.~\ref{fig:norot} we study the trace of the
matrix, $\tr(D)$, the first component $D_{xx}$ of which we subtracted half the
trace, and the off-diagonal element $D_{xy}$, symmetrised. The
component~$D_{yy}-\tr(D)/2$ is not plotted as it is simply the negative of
$D_{xx}-\tr(D)/2$. Only $\tr(D)$ still has hexagonal symmetry. The two bottom
panels show (skew) symmetry with respect to mirroring about the $x$-axis and
the $y$-axis. We found the same skew symmetry in the Hessian matrix,
Eq.~\eqref{H:xy}. The special choice of component combinations in
Fig.~\ref{fig:norot} is motivated by the continuum limit which we discuss now.
% >>>
\subsubsection{Hexagonal continuum limit}% <<<
\label{sec:hexcontinuum}

The matrix~$D_{\alpha\beta}(\bfn)$ is the inverse of the Green function of
static elasticity and is thus expected to be related to the corresponding
differential operator -- or rather its hexagonal discretisation. Continuum
elastic theory of pre-stressed materials gives the differential operator
(Sec.~3 of Ref.~\citenum{Wallace70})
\begin{equation}
  \label{calD}
  \calD_{\alpha\sigma} = -A_{\alpha\beta\sigma\tau}\partial_\beta\partial_\tau,
\end{equation}
where we have to remember that the two elastic tensors $A$~and $C$ coincide
only in systems whose reference state is stress-free. The hard-sphere crystal
in our numerical simulation is compressed by the periodic boundary conditions,
leading to the isotropic stress $T_{\beta\tau}=- P\delta_{\beta\tau}$ with the
pressure~$P$. Consequently, the two tensors $A$~and $C$ are different.
Assembling Eqs.~\eqref{C}, \eqref{AandC}, we find the differential operator to
be
\begin{equation}
  \calD_{\alpha\beta} = - (\mu-P)\delta_{\alpha\beta}\Laplace - (\lambda+\mu)\partial_\alpha\partial_\beta.
\end{equation}
For these second derivatives, we can produce simple discretisations in terms of
finite differences on nearest neighbours. In particular, the combinations of
matrix elements used in Fig.~\ref{fig:norot} are simple combinations of second
derivatives,\footnote{These stencils are obtained by writing the operator in
question as $\bfnabla^T M\bfnabla$, with a corresponding matrix $M$ (for
example, $M_{ij}=\delta_{ij}$ for $\bfnabla^2$). The matrix is then decomposed
into a suitable sum of terms $\bfe_i\bfe_j^T$, where the $\bfe_i$ are the six
hexagonal unit vectors obtained from rotation by $60^\circ$. The result is thus
composed of terms such as $\bfe_i^T\bfnabla$, which are directed derivatives
between the center and a first neighbour. Each of these terms is thus
immediately translated into a finite difference between them, resulting in the
weights noted in the stencils of Eqs.~\eqref{stencils}.}
\begingroup
\allowdisplaybreaks
\begin{subequations}
\label{stencils}
\begin{align}
  \tr\calD &\propto -\Laplace = \frac{1}{3d_0^2}\parbox{17mm}{\includegraphics{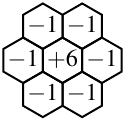}}\\
  \calD_{xx}-\frac{\tr\calD}{2} &\propto \partial_y^2 - \partial_x^2 = \frac{1}{3d_0^2}\parbox{17mm}{\includegraphics{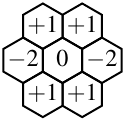}}\\
  \frac{\calD_{xy}+\calD_{yx}}{2} &\propto -\frac{\partial_x\partial_y+\partial_y\partial_x}{2} = \frac{1}{\sqrt{3}d_0^2}\parbox{17mm}{\includegraphics{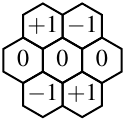}}
\end{align}
\end{subequations}
\endgroup
We see that these stencils are qualitatively reproduced in the central parts of
the panels in Fig.~\ref{fig:norot}. However, Fig.~\ref{fig:norot} shows nonzero
interactions also beyond the first layer of neighbours. Another difference is
that for $\tr(D)$ we find a middle value which is $-5.6$ times the values found
on the first neighbours (instead of~$-6$). When comparing the values of the
first neighbours among themselves, the stencils are again well reproduced, with
variations as small as~$\sim10^{-5}$ in all three panels of
Fig.~\ref{fig:norot}. The skew symmetry of the operator $\partial_x\partial_y$
is found again in all values of $D_{xy}(\bfn)$. This symmetry imposes the
horizontal and the vertical white lines (zeros) in the third panel of
Fig.~\ref{fig:norot}.

\begin{figure}%
  \centering
  \includegraphics{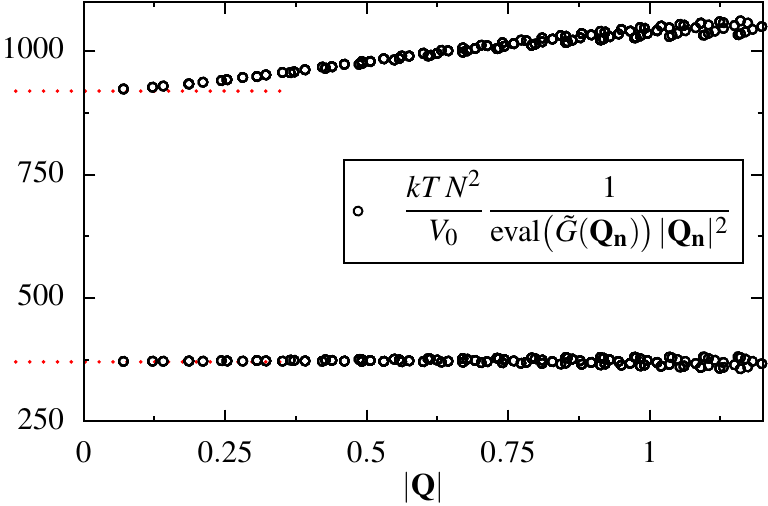}%
  \caption{Dispersion curves used to obtain the elastic moduli in the
  limit~$|\bfQ|\to0$, here for $L=100$. The dotted lines indicate values from a
  different numerical method working at~$|\bfQ|=0$.}%
  \label{fig:dispersion}%
\end{figure}

The matrix $D_{\alpha\beta}(\bfn)$ encodes the full dispersion curves that
are required to determine the spectral properties of the fluctuations. In
particular, one can extract the elastic tensor~$A$ from the long-wavelength
limit. The usual way to do this calculation is to perform a discrete periodic
(fast) Fourier transform on the matrix~$G_{\alpha\beta}(\bfi,\bfj)$ and to
observe that the translation invariance renders the result block-diagonal. For
each reciprocal vector~$\bfQ_\bfm$ we have one Fourier transformed
$d{\times}d$~matrix
\begin{align}
  \label{defFT}
  % notes p. 3661
  \tilG_{\alpha\beta}(\bfm)
   &:= N\sum_\bfn e^{-i\bfQ_\bfm\cdot\bfR_\bfn} G_{\alpha\beta}(\bfn) \\
    &= \Bigl\langle \overline{\tilu_\alpha(\bfm)} \tilu_\beta(\bfm)\Bigr\rangle.
\end{align}
The overline denotes a complex conjugate.
To leading order, this matrix scales as~$|\bfQ_\bfm|^{-2}$, such that we can
extract the elastic tensor as the long-wavelength prefactor to this scaling,
\begin{equation}
  \label{limitG}
  A_{\alpha\sigma\beta\tau}\hatQ_\sigma\hatQ_\tau
   = \frac{kT\,N^2}{V} \lim_{|\bfQ_\bfm|\to0} \frac{\bigl[\tilG(\bfm)^{-1}\bigr]_{\alpha\beta}}{|\bfQ_\bfm|^2}.
\end{equation}
In practice, rather the eigenvalues of the inverse matrix are plotted, see
Fig.~\ref{fig:dispersion}. The long-wavelength limit is then done by
extrapolation by eye. The upper curve corresponds to the longitudinal waves,
converging to $A_{xxxx} = (\mu-P)+(\lambda+\mu)$ in the limit $|\bfQ|\to0$. The
lower curve gives the transverse value $A_{xyxy}=\mu-P$.%
\begin{figure}%
  \centering
  \includegraphics{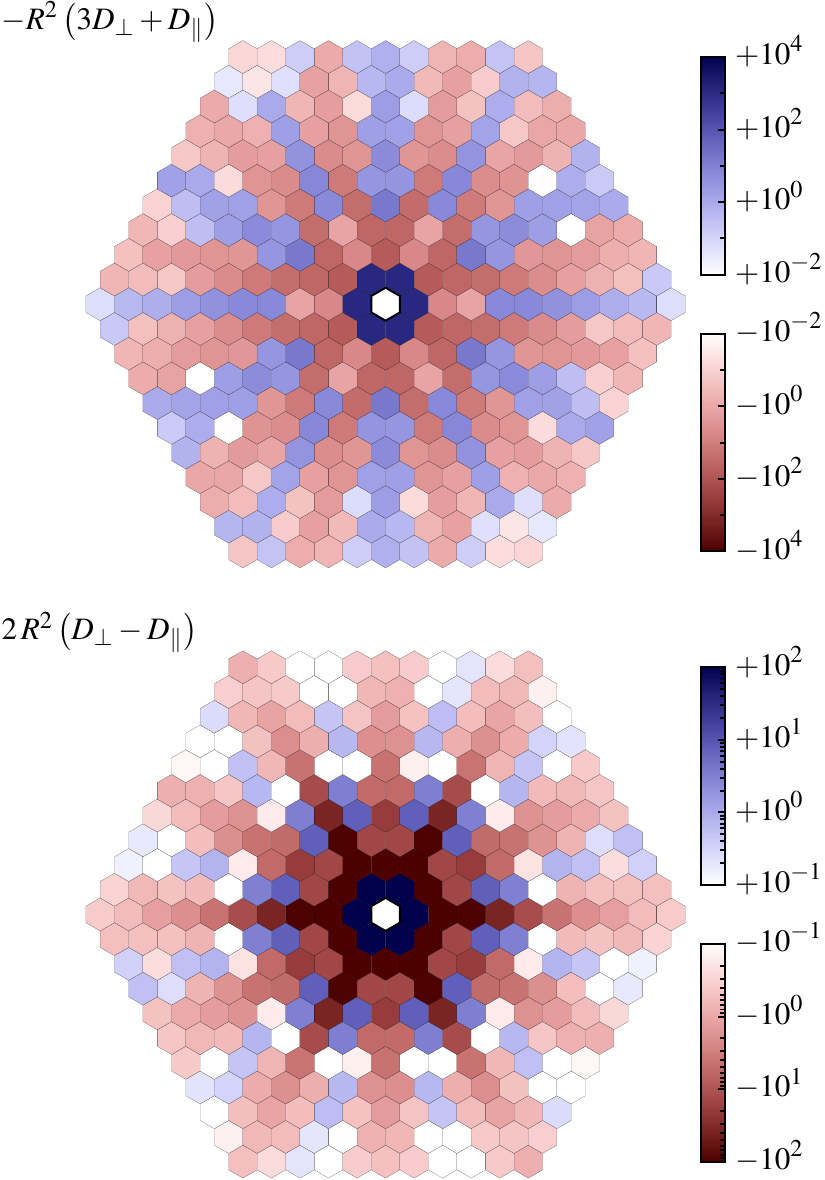}%
  \caption{The summands of Eqs.~\eqref{muP} and \eqref{mulam}.}%
  \label{fig:lame1}%
\end{figure}%
\begin{figure}%
  \centering
  \includegraphics{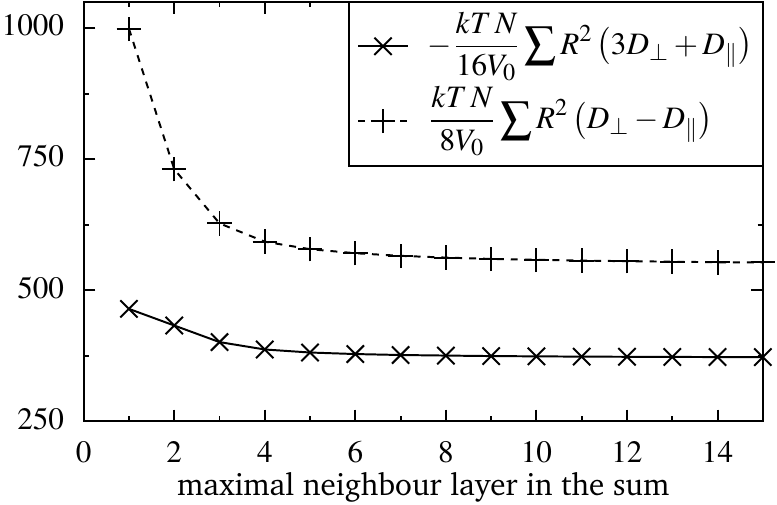}%
  \caption{Convergence of the sums in Eqs.~\eqref{muP} and \eqref{mulam}.}%
  \label{fig:lame2}%
\end{figure}%

The same limit can be done using the matrix~$D$ instead of~$G$. The block
structure helps to invert the large matrix, the inverse is again
block-diagonal. The $d{\times}d$~blocks are simply the Fourier transforms of the
matrices~$D_{\alpha\beta}(\bfn)$: $[\tilG(\bfm)^{-1}]_{\alpha\beta} =
\tilD_{\alpha\beta}(\bfm)/N^2$. For the long-wavelength limit we expand the
exponential in the definition of the Fourier transformation~\eqref{defFT}
around $\bfQ_\bfm=0$. The sum over the constant term vanishes, $\sum_\bfn
D_{\alpha\beta}(\bfn)=0$. The linear term vanishes due to symmetry, such that we obtain the
quadratic term as the leading one. This conveniently matches with what we
require for the limit in Eq.~\eqref{limitG}. Factorising the unit reciprocal
vectors, we obtain for the elastic tensor
\begin{equation}
  \label{limitD}
  A_{\alpha\sigma\beta\tau}
   = -\frac{1}{2} \frac{kT\,N}{V} \sum_{\bfn} R_{\bfn\sigma} R_{\bfn\tau} D_{\alpha\beta}(\bfn).
\end{equation}
To determine the Lamé coefficients, we can either choose individual components such
as $A_{xxxx}$ above, or we can do a ``hexagonal average'', which amounts to
reducing the rank-4 tensor to scalars, for example with
$\delta_{\alpha\beta}\delta_{\sigma\tau}$ and with
$\delta_{\alpha\sigma}\delta_{\beta\tau}$. This gives a $2{\times}2$ system of
equations for $(\mu-P)$ and $(\lambda+\mu)$, which when solved becomes
\begin{align}
\label{muP}
  \mu-P &= -\frac{1}{16}\frac{kT\,N}{V} \sum_{\bfn} R_{\bfn}^2 \bigl(3D_\perp(\bfn) + D_\parallel(\bfn)\bigr) \\
\label{mulam}
  \lambda+\mu &= \frac{1}{8}\frac{kT\,N}{V} \sum_{\bfn} R_{\bfn}^2 \bigl(D_\perp(\bfn) - D_\parallel(\bfn)\bigr)
\end{align}
The summands are displayed in Fig.~\ref{fig:lame1}. They decrease with
increasing distance, as expected, and their sums converge. We visualise the
convergence in Fig.~\ref{fig:lame2} where we added up the summands layer by
layer. The final values are $\mu-P=373$ and $\lambda+\mu=553$.

In order to extract the Lamé coefficients, we need also the value of the
pressure~$P$. During simulations, we tracked the average flux of linear
momentum, which is a mechanical definition of the pressure tensor. We further
measured the response of this tensor to small deformations of the simulation
box, which presents another (faster converging) way to calculate the elastic
moduli. Notice that this method corresponds directly to ``$|\bfQ|=0$'', no
limit has to be taken. We obtain the values~\footnote{These values are
compatible with those given in Ref.~\citenum{WojTreBraKow03}. We applied the
box-deformation method also to the value $\phi=0.863714$ which is given in that
reference, and our implementation reproduces exactly the given values for the
elastic moduli. Concerning the numerical values of elastic moduli, there was a
disagreement in the literature between Ref.~\citenum{binder} and
Ref.~\citenum{WojTreBraKow03}, see also subsequent publications. Given that we
implemented both the fluctuation method and two deformation methods (in a
second one the spheres are deformed instead of the box) which all give the same
result, we think that we can resolve the debate in favour of
Ref.~\citenum{WojTreBraKow03}.}
\begin{equation}
  \label{Pmulamb}
  P = 34.4, \quad \mu=405, \quad \lambda=143.5\, .
\end{equation}
They are used in Fig.~\ref{fig:dispersion} to check the consistency of the
different methods. Also the values extracted from Eqs.~\eqref{muP} and
\eqref{mulam} are reasonably close.

% >>>
\subsection{Three-dimensional crystal}% <<<
\begin{figure}%
  \centering
  \includegraphics{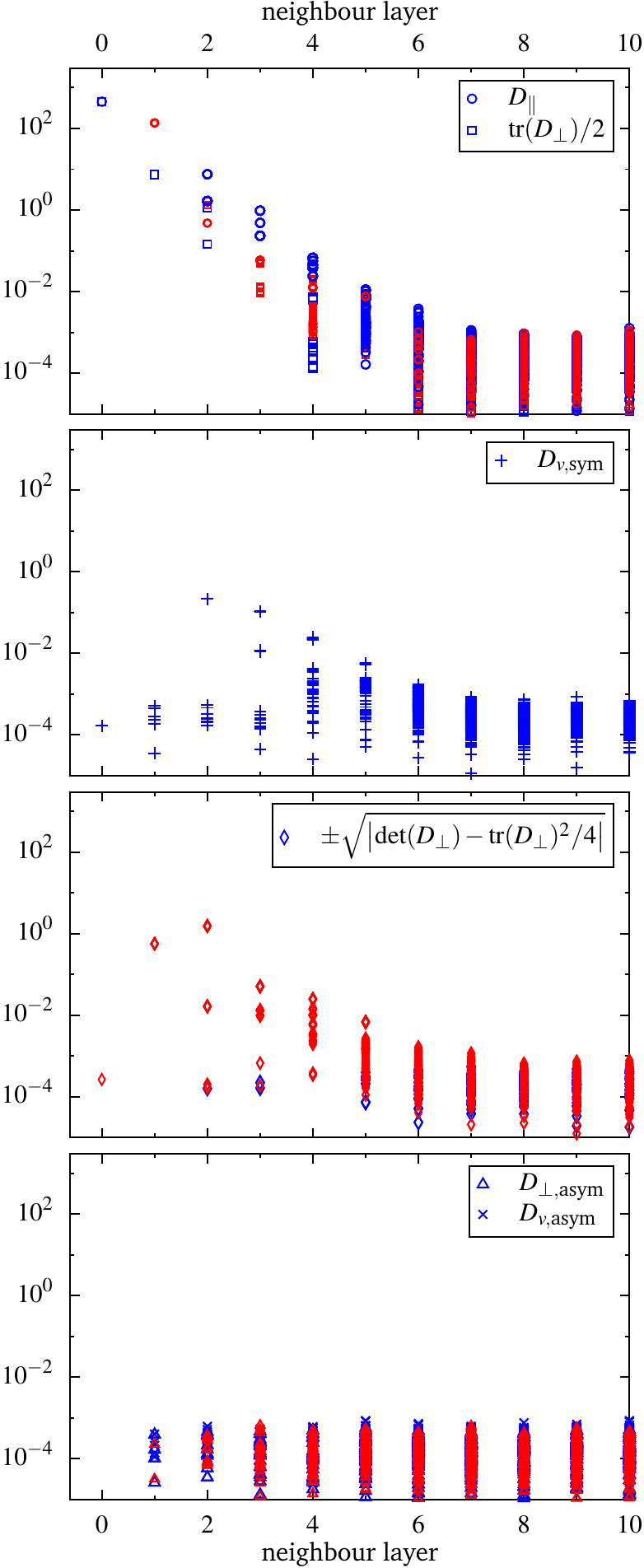}%
  \caption{Effective interactions in a FCC crystal of $N=15^3$ hard spheres.
  Larger blue symbols are positive values, smaller red symbols are negative
  values. $K=3N\times 6.8{\times}10^4$ recordings, obtained from
  424~independent trajectories; volume fraction~$\phi=0.57$. These data
  required around 230\,000 CPU core hours.}%
  \label{fig:3D}%
\end{figure}%
We also performed simulations in three dimensions, collecting high statistics
data on a system of dimensions $L=15$. Here, the result is similar to two
dimensions, the effective interaction does not vanish beyond the layer of
nearest neighbours but shows a rapid decrease. We adopt the same route of a
rotated frame of reference to plot in Fig.~\ref{fig:3D} the matrix elements of
the $3{\times}3$~matrices $D_{\alpha\beta}(\bfn)$. We choose an orthonormal
basis~$(\hat\bfR_\bfn,{\hat\bfo}_1,{\hat\bfo}_2)$, in which the matrix comprises
the following blocks:
\begin{equation}
  \left(\begin{array}{ccc}
   D_\parallel & \bfv^T \\
   \bfw & D_\perp
  \end{array}\right)
\end{equation}
As before, $D_\parallel=\hat\bfR_\bfn^T D\hat\bfR_\bfn$, but $D_\perp$ is
a $2{\times}2$~matrix, and $\bfv,\bfw$ are two-dimensional vectors. As the
basis vectors ${\hat\bfo}_1,{\hat\bfo}_2$ can be chosen with an arbitrary
rotation about the vector~$\hat\bfR_\bfn$, we are only interested in invariants
of $D_\perp,\bfv,\bfw$ under this rotation. For the matrix~$D_\perp$ we thus
plot in Fig.~\ref{fig:3D} the trace of $D_\perp$, its anti-symmetrised
off-diagonals $D_{\perp,\text{asym}}$ and some notion of its determinant. For
$\bfv,\bfw$ we plot the modulus after (anti)-symmetrising,
$D_{v,\text{sym}}:=\|\bfv+\bfw\|/2$, $D_{v,\text{asym}}:=\|\bfv-\bfw\|/2$. The
lowest panel in Fig.~\ref{fig:3D} allows one to estimate the noise level for the
given statistics.

If again one insists on an algebraic fit $D(\bfn)\propto\|\bfR/d_0\|^\alpha$,
one finds $\alpha\approx-8\pm1$. In three dimension the data display less
scatter when plotted in terms of neighbour layer, rather than separation. This
is a non-trivial geometrical effect which is already visible in two dimensions:
In the first panel of Fig.~\ref{fig:dist_rot}, the values on the diagonals of
the hexagon are larger than the values at the same distance (even larger than
those on the same neighbour level). This creates within every neighbour level
(or distance) a tendency which is \emph{opposed} to the general trend.
% >>>

\section{Discussion and conclusions}
\label{sec:discussion}
% <<<
In section~\ref{sec:numerics} we have presented a number of properties, some of
which are compatible, others are incompatible with the expectation elaborated
from the central model in section~\ref{sec:theory}. Let us now scrutinise these
properties and see to what extend we can understand the differences.

\paragraph{Elastic moduli:} Let us start with the elastic moduli and the Cauchy
relation. The values we found for the Lamé coefficients in Eqs.~\eqref{Pmulamb}
show that the two-dimensional hexagonal crystal is far from being ``Cauchy'',
$\mu$~being nearly three times as large as~$\lambda$, where the Cauchy
relations would require them to be equal. This finding is neither
new\cite{WojTreBraKow03} nor surprising, as shows the calculation
by~\citet{hoover}, which we recapitulated in Sec.~\ref{sec:hoover}. One
concludes that the non-thermal central force model is inadequate for
understanding the elastic moduli of the hard-sphere system, that the kinetic
term and probably more important the fluctuations term are not small, and
finally that we can indeed expect the hard-sphere system to be dominated by
thermal effects and to see important deviations also in the interaction matrix.

%If we blindly apply Eq.~\eqref{barmu} to the numerical data, using
%$R^2(D_\perp-D_\parallel)$ as the summand, then we simply calculate the average
%value $(\mu+\lambda)/2$.

\paragraph{Pressure:} As a side remark, we find it an interesting question
whether one can determine the pressure in a system by measuring displacements
and their correlations.\cite{DanielBonn} Equation~\eqref{barT} shows that this
is indeed possible in the non-thermal model, where the pressure is minus a
diagonal term of the stress in Eq.~\eqref{barT}, or
$\barP=-\frac{N}{4V}\sum_\bfn R_\bfn^2 g'_\bfn\bigl(R_\bfn^2/2\bigr)$. The
values of~$g_\bfn^\prime$ are translated back to $H_\perp$ using
Eq.~\eqref{Hrot:ortho}. Does the same work also in the thermal case, using
$D_\perp$ instead of $H_\perp$? When we use the numerical data for $D_\perp$,
shown in Fig.~\ref{fig:dist_rot}, then we find the value~$-96$ for this
effective pressure -- which is clearly unacceptable, even the sign is wrong. In
fact, a close look on Eqs.~\eqref{muP} and \eqref{mulam} reveals that the sum
actually calculates $P+(\lambda-\mu)/2$ and equals the pressure only to the
extent that the Cauchy relation is satisfied.

\paragraph{Locality of the interaction:} Independent of all questions of how
the data compare to a model, we found that the correlation matrix strongly
depends on the system size, whereas its inverse does not (Fig.~\ref{fig:size}).
Our intuition, in which we called the correlations a ``Green function'' and its
inverse a ``differential operator'' heavily relies on a clear-cut separation of
size dependence. We find that the differential operator ($D$) is indeed
``local'', but the precise notion of this locality is different in one and in
higher dimensions. In one dimension we find the interaction matrix to be
restricted to the two nearest neighbours, as expected from a Laplace operator.
In two and three dimensions the results are much more interesting: Despite the
true interactions taking place to nearest neighbours, we find longer-ranged
effective interactions from the analysis of displacement correlations. These
interactions are small compared to the ones between nearest neighbours, but
they are not exactly zero and can be distinguished from noise. It remains open
what functional form the decrease in Figs.~\ref{fig:2D} and \ref{fig:3D}
follows, and whether an algebraic form would still allow to reasonably call the
operator ``local.''

\paragraph{Theoretical interpretation of the dynamic matrix as a self-energy:}
The presented results open the question of a better theoretical
interpretation than the one in Sec.~\ref{sec:local}. In particular one would
like to {\sl predict} how the effective interactions decay with distance. It is
clear that in a field-theory formulation of a perturbed Gaussian theory the
effective interactions $D$~correspond to the self-energy in a Dyson equation.
However, without an explicit perturbation scheme for hard sphere systems it is
difficult to turn this remark into an explicit calculation.

\paragraph{Nonlinearities in the interaction potential:} Getting back to the
central-force model of Sec.~\ref{sec:local}, it does not favour local or
nonlocal interactions in its general form. If we want more information, we need
to specify the functions~$g_{\bfi\bfj}$. One choice is to take the true
hard-sphere interactions, the pair potential being either zero or infinity.
Around the reference state, the potential is zero, together with its first and
second derivative. Therefore, the Hessian matrix vanishes, and also the stress
and elastic tensors consist only in the kinetic terms. Here, the model is
clearly out of its applicability.

We learn from this remark that the model equations do not necessarily serve to
infer the true interaction between particles from given data. But, what lesson
do we want to learn from the model? Our aim was to contrast thermal with respect to
non-thermal effects. Unfortunately, in the hard-sphere simulations, together
with thermal effects, we introduced strong nonlinearities in the pair
potential. Now we have a hard time to separate their influence from the thermal
effects. (At least in the elastic constants, the nonlinearities play no role,
which is reassuring.)

Another convenient choice for the functions~$g_{\bfi\bfj}$ is a network of
linear springs which have zero length at rest. When such a network is stretched, the
particles may be found on a hexagonal reference lattice. In this system the
Hessian matrix is a constant, and all statistical averages are Gaussian
integrals which can be solved analytically via Wick's theorem. The effective
interactions are indeed found to be given by the Hessian matrix, $D=H/kT$,
which in turn is given by the connections of particles. If we choose to have
identical springs between nearest neighbours only, then also the matrix~$D$ is
restricted to nearest neighbours. The same applies to second-nearest
neighbours, and so on. One might think that this linear system will help to
differentiate between thermal and nonlinear effects. However, when one
evaluates the averages in Eq.~\eqref{hoover:elast}, which contains all thermal
effects, one finds that the elastic tensor is zero.\footnote{This does not mean
that the sound waves in this system have vanishing sound velocity.
Sound velocities as well as phonons are encoded in the tensor~$A$ rather
than~$C$, and only the latter vanishes. The tensor~$A$ has additional
contributions which are proportional to the stress of the reference state, see
Eq.~\eqref{AandC}.} The zero-length spring network is thus not good enough to
separate nonlinear from thermal effects.

If we insist on central pair potentials, the zero-length spring network is the
only linear one we know of. Of course, other arbitrary linear models can be
created by choosing a potential of the form~\eqref{DeltaPhi}, i.\,e.~exactly
quadratic in the displacements -- one can even take the measured matrix~$D$ as
the matrix in this expression, which amounts to the \emph{shadow system} of
Ref.~\citenum{silke}. However, doing so one relaxes the constraint of the
central potential and includes knowledge about the reference positions in the
potential. Moreover, by construction the shadow system has the same
displacement correlations, the same interaction matrix, and the same elastic
constants as the original system. Only higher-order correlations differ. We are
thus back to the question what do we want to learn from comparing with a model?
Should it contain information about the reference positions, thus about the
spontaneous symmetry breaking or not?

Another possible approach to a model would consist in extracting the ``best''
central-force model from the numerical data. Wanted is a discrete
function~$g_\bfn$ of which the first and second derivatives are given by
$g_\bfn'=-D_\perp(\bfn)$ and $g_\bfn^{\prime\prime} =
\bigl(D_\perp(\bfn)-D_\parallel(\bfn)\bigr)/R_\bfn^2$, respectively -- in the
hexagonal case, in analogy to Eqs.~\eqref{Hrot:parall} and \eqref{Hrot:ortho}.
The first derivative can be read off Fig.~\ref{fig:dist_rot}, and the second
one is given up to a factor~$R_\bfn^4$ in the lower panel of
Fig.~\ref{fig:lame1}. We did not try to perform this extraction and do not know
whether the problem has a unique solution or approximation methods are
required.

\paragraph{Non-central forces:} The main lesson we can learn from the
comparison with the model is probably that the effective interactions contained
in~$D$ are non-central: It is striking that $H_\text{sym}(\bfn)$ is identically
zero for all~$\bfn$, whereas $D_\text{sym}(\bfn)$ has nonzero values. \emph{The
thermal interactions are thus genuinely non-central.} We note that the
off-diagonal term $D_\text{sym}(\bfn)$ is smaller in amplitude than the terms
on the diagonal, $D_\parallel$ and $D_\perp$, so that the non-central nature
can be considered a correction to the dominant terms.

% >>>

%\bibliographystyle{rsc}
%\bibliography{md}
\providecommand*{\mcitethebibliography}{\thebibliography}
\csname @ifundefined\endcsname{endmcitethebibliography}
{\let\endmcitethebibliography\endthebibliography}{}

\end{document}